\documentstyle[12pt,aaspp4,epsf]{article}
 
\newdimen\hheight
\newdimen\hwidth
\newdimen\theight
\newdimen\twidth
\newdimen\colspace

\hheight=\textheight \divide\hheight by 2
\hwidth=\textwidth \divide\hwidth by 2 \advance\hwidth-2\colspace
\theight=\textheight \divide\theight by 3
\twidth=\textwidth \divide\twidth by 3 \advance\twidth-2\colspace

\def\sech{\mathop{\rm sech}\nolimits}

\begin{document}

\title{An adaptive algorithm for n-body field expansions}

\author{Martin D. Weinberg}

\affil{Department of Physics \& Astronomy \\
  University of Massachusetts, Amherst, MA 01003-4525}

\abstract{An expansion of a density field or particle distribution in
  basis functions which solve the Poisson equation both provides an
  easily parallelized n-body force algorithm and simplifies
  perturbation theories.  The expansion converges quickly and provides
  the highest computational advantage if the lowest-order
  potential-density pair in the basis looks like the unperturbed
  galaxy or stellar system.  Unfortunately, there are only a handful
  of such basis in the literature which limits this advantage.  This
  paper presents an algorithm for deriving these bases to match a wide
  variety of galaxy models.  The method is based on efficient
  numerical solution of the Sturm-Liouville equation and can be used
  for any geometry with a separable Laplacian.
  
  Two cases are described in detail.  First for the spherical case,
  the lowest order basis function pair may be chosen to be exactly
  that of the underlying model.  The profile may be cuspy or have a
  core and truncated or of infinite extent.  Secondly, the method
  yields a three-dimensional cylindrical basis appropriate for
  studying galaxian disks.  In this case, the vertical and radial
  bases are coupled; the lowest order radial part of the basis
  function can be chosen to match the underlying profile only in the
  disk plane.  Practically, this basis is still a very good match to
  the overall disk profile and converges in a small number of terms.
  }


\section{Introduction} \label{sec:intro}

The basis function n-body force solver is optimal for studying the
global response of galaxies to perturbations or stability (Earn \&
Sellwood 1995\nocite{EaSe:95}).  This technique was developed for
astrophysical problems by Clutton-Brock (1972,
1973\nocite{Clut:72,Clut:73}), Kalnajs (1976\nocite{Kaln:76a}),
Fridman \& Polyachenko (1984\nocite{FrPo:84}) and more recently by
Hernquist \& Ostriker (1992\nocite{HeOs:92}) who dubbed it the {\it
  self-consistent field} (SCF) method.  Orthogonal function expansions
are attractive Poisson equation solvers for two reasons: 1) the
expansions can be chosen to filter the structure over an interesting
range of scales and simultaneously suppress small-scale noise; and 2)
the algorithm is computationally efficient, scaling linearly with the
number of particles.  Mathematically, this entire class of algorithms
relies on the general properties of the Sturm-Liouville equation (SLE)
of which the Poisson equation is a particular case.  This same
approach is common in perturbation theories and so facilitates direct
comparison between n-body simulation and linear perturbation theory.
In addition, this approach is straightforward to parallelize (e.g.
Hernquist, Sigurdsson \& Bryan 1995\nocite{HeSB:95}); we find the
algorithm scales linearly with the number of processors with low
overhead.  If the basis set resembles the equilibrium galaxy, most of
the computational work is concentrated on resolving the perturbation
rather than the equilibrium.

This last point is also a disadvantage of this technique in
applications to date.  If the equilibrium does not look like the basis
set, the technique becomes less efficient and noisy because the
expansion series must be sufficiently long to represent the
equilibrium even without the perturbation.  This paper describes a
general method based on a numerical construction of orthogonal bases
which remedies this situation.  Solutions to the fundamental equation,
the Sturm-Liouville equation, are well-understood and well-behaved.  A
number of recently published algorithms take advantage of the special
properties of this differential equation to yield high-accuracy
solutions with low computational work.  Harnessing these developments
to our needs leads to an algorithm for computing orthogonal bases
whose lowest-order function matches any given {\it any} regular
equilibrium; spherical and three-dimensional cylindrical solutions are
described in detail here.  The basic algorithm will be described in
\S\ref{sec:SL}.

For the spherical case, the proposed algorithm is competitive in
performance with evaluation by recursion relation used for the
published bases cited above and has reproduced them with high accuracy
as a check.  The cylindrical basis is a bit more cumbersome: one may
rely on the same numerical solution to tailor the basis in the radial
or vertical direction but not both simultaneously.  Here, I choose to
derive the radial basis numerically.  The lowest-order radial basis
functions then take the form $f(r)\exp(\pm ikz)$.\footnote{Bases
  resulting the other choice has been explored by Earn
  (1996\nocite{Earn:96}) using a different approach.}  These may then
be adapted to the background by principal component analysis.
Although more cumbersome to implement and more time consuming to
execute than the spherical case, it is still fast relative to
non-expansion-based solvers.  The details of the cylindrical basis are
given in \S\ref{sec:PCA}.

\section{The algorithm} \label{sec:SL}

\subsection{Motivation}
\label{sec:motiv}

Here, I will explicitly describe the spherical and three-dimensional
disk cases but all others are analogously derived with little change.

The Poisson equation separates in any conic coordinate system.
Choice of separation constants gives a differential equation in the
SLE form for each dimension.  The simplest solution employs the
eigenfunctions of the Laplacian directly.  For example, consider an
expansion in spherical polar coordinates.  Assuming that the density
is proportional to the potential, the solution to Poisson's equation
takes the form of an eigenfunction of the Laplacian:
\begin{equation}
  \label{eq:efunclap}
  {d^2R(r)\over dr^2} + {2\over r} {dR(r)\over dr} - {l(l+1)\over r^2}
  R(r) = 4\pi G\lambda R(r).
\end{equation}
The well-known full solution is the product of spherical harmonics in
$\theta$ and $\phi$ and Bessel functions in $r$.  For a finite-radius
mass distribution with an inner core, the inner boundary condition is
the usual $dR/dr|_0 = 0$ and the multipole expansion provides the
outgoing boundary condition:
\begin{eqnarray}
  \label{eq:bndry}
  \left.{dR(r)\over dr}\right|_{r_t} = \left.-(l+1) {R(r)\over r}\right|_{r_t},
\end{eqnarray}
where $r_t$ is the outer edge of the profile.  Using these boundary
conditions and the orthogonality relation of the Bessel functions
leads to the following potential and density pair:
\begin{eqnarray}
  \label{eq:orthopr}
  p_n^{lm}(r) &=& {1\over  a_n^{lm} |J_{l+1/2}(a_n^{lm})|} \sqrt{2\over r}
  J_{l+1/2}(a_n^{lm} r/r_t), \nonumber \\
  d_n^{lm}(r) &=& {a_n^{lm} \over r_t^2 |J_{l+1/2}(a_n^{lm})|} \sqrt{2\over r}
  J_{l+1/2}(a_n^{lm} r/r_t),
\end{eqnarray}
where $a_n^{lm}$ is the $n^{th}$ zero of $J_{l-1/2}$ and $r_t$ is the
outer edge of the profile (Fridman \& Polyachenko
1984\nocite{FrPo:84}).  The functions $p_n^{lm}$ and $d_n^{lm}$ have
the following inner product:
\begin{equation}
  \label{eq:scalar}
  \int^\infty_0 dr\, r^2\, p_n^{lm}(r) d_{n^\prime}^{lm\,\ast}(r) =
  -\delta_{n\,{n^\prime}}.
\end{equation}
Properties of solutions to the SLE ensure that this expansion set is
complete.  Therefore given a density distribution, the gravitational
potential and force can be found directly by expansion.  The set
$(p_n, d_n)$ are often called {\it biorthogonal}.  A similar expansion
obtains for cylindrical polar coordinates.

This straightforward approach has flaws.  Bessel functions do not look
like galaxian profiles and therefore accuracy demands high-order
expansions.  The required number of functions increases for extended
profiles since Bessel functions are only orthogonal over a finite
domain.  To get around this, one may map the radial coordinate from
the semi-infinite real axis to a finite segment.  Appropriate choice
of this transformation leads to new sets of biorthogonal functions in
both the spherical (Clutton-Brock 1973, Hernquist \& Ostriker
1992\nocite{Clut:73,HeOs:92}) and two-dimensional (Clutton-Brock 1972,
Kalnajs 1976) and three-dimensional (Earn
1996\nocite{Clut:72,Kaln:76a,Earn:96}) cylindrical cases.  This small
number of choices results in a mismatch between the lowest order basis
functions and equilibrium profile. A poor fit between the basis and
the underlying density profile is a source of noise in the force field
which leads to relaxation (cf. Weinberg 1997\nocite{Wein:97b}).  This
is the general situation unless one's galaxy fortuitously coincides
with particular sets of orthogonal polynomials or functions
analytically derived from exact solutions of the Poisson equation.

The solution proposed here is a numerical solution of the SLE using
recently developed and published techniques (Marletta \& Pryce 1991,
Pruess \& Fulton 1993, see Pryce 1993 for a
review\nocite{MaPr:91,PrFu:93,Pryc:93}).  This allows adaptive
construction of an expansion basis which matches the underlying
density profile exactly and thereby removes one of the major
limitations of this approach.  The details are described in the next
two sections.

Alternative solutions to the mismatch problem have been described by
Allen, Palmer \& Papaloizou (1990\nocite{AlPP:90}) and Saha
(1993\nocite{Saha:93}).  Both of these methods in their general form
rely on the orthogonalization of a covering but non-orthogonal basis.
There are two advantages to the approach developed here.  First, the
background profile is represented in one basis function with
potentially rapid convergence in the perturbation.  The basis
evaluation is easily incorporated into existing SCF codes.  Second,
the same biorthogonal series may be used in linear perturbation
analyses (e.g. Kalnajs 1976\nocite{Kaln:76a}, Fridman \& Polyachenko
1984\nocite{FrPo:84}, Weinberg 1990\nocite{Wein:90}) and coefficients
directly compared with n-body simulation.  This development was
motivated for precisely this reason and will underlie future inquiry.

\subsection{Reduction of the Poisson equation to Sturm-Liouville form}
\label{sec:deriv}

We present the cylindrical polar case here to be explicit but again
the others are analogous.  The Laplace equation separates into the
following three equations for a potential of the form $\Psi({\bf r}) =
R(r)Z(z)\Theta(\theta)$:
\begin{eqnarray}
  \label{eq:laplace}
  {1\over r}{d\over dr}r{d\over dr} R(r) -\left(k^2 +
    {m^2\over r^2}\right) R(r) &=& 0 \nonumber \\
  {d^2\over dz^2}Z(z) + k^2 Z(z) &=& 0 \nonumber \\
  {d^2\over d\theta^2}\Theta(\theta) + m^2 \Theta(\theta) &=& 0.
\end{eqnarray}
Following the authors cited in \ref{sec:motiv}, we can look for a
solution to the Poisson equation whose potential and density have the
form
\begin{eqnarray}
  \label{eq:pairs}
  \Psi(r, z, \theta) &=& \Psi_o(r) u(r) Z(z) \Theta(\theta) \nonumber \\
  \rho(r, z, \theta) &=& \rho_o(r) u(r) Z(z) \Theta(\theta).
\end{eqnarray}
The Poisson equation then takes the form
\begin{equation}
  \label{eq:poisson}
  {1\over r}{d\over dr}r{d\over dr} \Psi_o(r) u(r) -\left(k^2 +
    {m^2\over r^2}\right) R(r) = 4\pi G \lambda \rho_o(r) u(r)
\end{equation}
together with second two of equation (\ref{eq:laplace}) above,
where $\lambda$ is an unknown constant.

The general form of the SLE is usually quoted as:
\begin{equation}
  \label{eq:SLE}
  -{d\over dx}\left(p(x){du\over dx}\right) + q(x)u = \lambda w(x) u
\end{equation}
where $p(x), w(x)>0$ over the domain of interest, $[a, b]$.  The
eigenfunctions are orthogonal (see Courant \& Hilbert
1953\nocite{CoHi:53} for extensive discussion) and may be normalized:
$\int_a^b dx\, w(x) u^2 = 1$.  Equation (\ref{eq:poisson}) is easily
rewritten in this form and one finds:
\begin{eqnarray}
  \label{eq:poisson2}
  {d\over dr}\left[ r\Psi_o^2(r) {d u(r)\over dr} \right] - 
  \left[ k^2 \Psi_o(r) + {m^2\over r^2}\Psi_o(r) -
    \nabla_r^2\Psi_o(r)\right] r \Psi_o(r) u(r) = \nonumber \\
  \qquad 4\pi G \lambda r\Psi_o(r) \rho_o(r) u(r)
\end{eqnarray}
where $\nabla_r$ denotes the radial part of the Laplacian operator.
The unknown constant $\lambda$ is the eigenvalue.  Comparing to the
standard SLE form, we have
\begin{eqnarray}
  \label{eq:sle1}
  p(r) &=& r\Psi_o^2(r), \\
  \label{eq:sle2}
  q(r) &=& \left[ k^2 \Psi_o(r) + {m^2\over r^2}\Psi_o(r) -
    \nabla_r^2\Psi_o(r)\right] r \Psi_o(r), \\
  \label{eq:sle3}
  w(r) &=& -4\pi G r\Psi_o(r)\rho_o(r).
\end{eqnarray}
These coefficient functions now provide the input to the standard
packaged SLE solvers either in tabular or subroutine form.  The
orthogonality condition for this case is
\begin{equation}
  \label{eq:ortho}
  -4\pi G \int^\infty_0 dr\, r\, \Psi_o(r)\rho_o(r) u(r)^2 = -4\pi
  G\int^\infty_0 dr\, r \,\Psi \rho = 1.
\end{equation}
In other words, equations (\ref{eq:pairs}) are potential-density
pairs.  It is convenient to define ${\tilde\rho}\equiv4\pi G\rho$ so
that the biorthogonality relation becomes $\int
dr\,r\,\Psi_r(r){\tilde\rho}_s(r)=-\delta_{rs}$.  Analogous
expressions obtain for the spherical polar case.  This development
does not require that $\Psi_o$ and $\rho_o$ solve the Poisson equation
but they must obey the appropriate boundary conditions at the center
and at the edge (which may be $r=\infty$).  If we choose $\Psi_o$ and
$\rho_o$ to be a solution of the Poisson equation then the lowest
eigenvalue is unity and the eigenfunction $u(r)$ is a constant
function.

\subsection{Numerical solution}
\label{sec:numerical}

For our problem, the SLE is well-conditioned and generally stable.
Solutions may be straightforwardly obtained by shooting methods and
standard ODE packages.  Here, I used the Pruess method (Pruess
1973\nocite{Prue:73}) as implemented by Pruess \& Fulton
(1993\nocite{PrFu:93}) with excellent success.  Rather than find an
approximate solution to the exact differential equation in the usual
way, this approach approximates the differential equation by a
piecewise continuous function---a discrete grid---and finds an exact
solution to the approximate problem.  The grid may be successively
refined to ensure convergence to the desired tolerance.  Additional
numerical analysis provides the optimal choice of grid over the domain
(which, again, may be infinite).  This choice of a non-uniform grid is
the numerical analog to transformation of the infinite interval to a
discrete segment which plays a defining role in Clutton-Brock's
approach.

The resulting numerical eigenfunctions must be tabulated for future
use.  By contrast, the orthogonal polynomial schemes yield explicit
recursion relations and this lack is the only practical disadvantage
to this approach.  On the other hand, the numerical SLE approach gives
us the flexibility to specify $\Psi_o$ and $\rho_o$ arbitrarily.  For
example, we may use the density profile from a previous n-body
simulation.

\section{Examples and comparisons}

\subsection{Spherical solutions for galaxian halos
  and spheroids}

\subsubsection{Method}

The boundary conditions must be appropriate for the problem at hand.
In the case of spherical symmetry, there is a boundary at $r=0$ and
$r=r_t$.  The inner boundary condition may be the traditional
$\Psi^\prime=0$ or that for a scale-free cusp.  The outer boundary
condition follows from the multipole expansion:
\begin{equation}
  \label{eq:sphbc}
  {d\Psi(r)\over dr} = -{l+1\over r}\Psi(r).
\end{equation}
We may have $r_t\rightarrow\infty$ in which case equation
(\ref{eq:sphbc}) applies in this limit.  Once the functions are
tabulated, the force algorithm proceeds as usual for an SCF code.
Given $\Phi_o$ and $\rho_o$, equations
(\ref{eq:sle1})--(\ref{eq:sle3}) define the eigenvalue problem for the
SLE.  For example, the Pruess \& Fulton code returns the
eigenfunctions $u(r)$ and the potential-density pairs follow from
equations (\ref{eq:pairs}).  The basis functions can be periodically
recomputed to adaptively fit an evolving distribution; we have not
implemented this for the spherical case here but see \S\ref{sec:PCA}
and Weinberg (1996\nocite{Wein:96}).

\subsubsection{Examples}

To test the spherical implementation, I assigned $\Psi_o$ and $\rho_o$
to the Hernquist model (Hernquist 1990\nocite{Hern:90}) and compared
the SLE solution with the analytic recursion relations (Hernquist \&
Ostriker 1992) for radial order $n\le16$ and $m\le2$.  Performance of
the spherical algorithm is well-documented so a comparison of
potential pairs suffices.  For $m=0$, the numerically determined
functions differed from the results of the recursion relation by one
part in $10^3$ near the center and one part in $10^6$ elsewhere.  This
difference is due to the extrapolation of the cusp at $r=0$.  Here,
the boundary condition for the cuspy profile fixes the asymptotic
value of ratio $\Psi_o^\prime/\Psi_o$ as $r\rightarrow0$.  For $m>0$
the differences are obtained to the specified tolerance (one part in
$10^6$ for these tests).  To recover the Clutton-Brock (1973) set, one
assigns $\Psi_o$ and $\rho_o$ according to the Plummer law; in this
case, differences between the SLE solution and recursion relations are
obtained for all $m$ to the desired tolerance. In all cases, the
orthogonality relation remains accurate and the potential density pair
is an accurate solution of the Poisson equation.

\begin{figure}[htb]
\begin{center}
  \mbox{\epsfxsize=4.0in\epsfbox{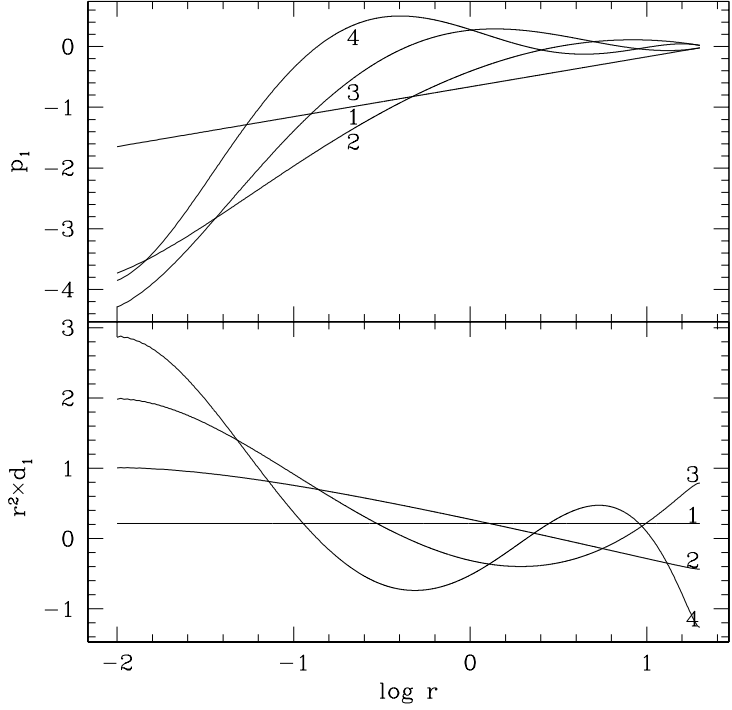}}
  \caption{Potential-density pairs for $l=m=0$ labeled by order,
    $n=1,\ldots4$ (upper and lower panels, resp.) whose lowest order
    member ($n=1$) is the singular isothermal sphere.  The density
    eigenfunctions are multiplied by $r^2$.}
  \label{fig:singiso}
\end{center}
\end{figure}

\begin{figure}[htb]
\begin{center}
  \mbox{\epsfxsize=4.0in\epsfbox{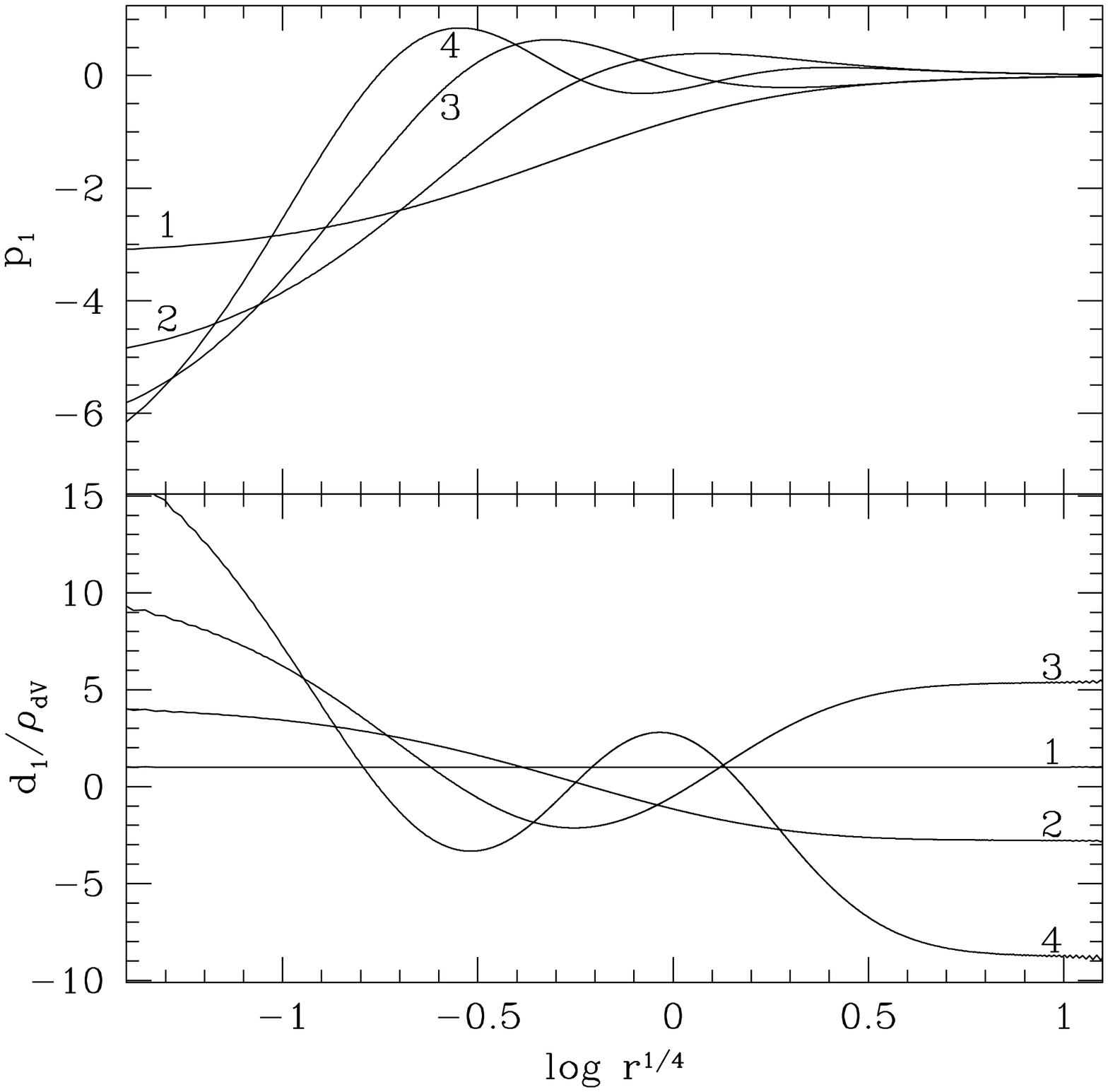}}
  \caption{Potential-density pairs for $l=m=0$ labeled by order,
    $n=1,\ldots4$ (upper and lower panels, resp.) whose lowest order
    member ($n=1$) is the spherical deprojection of the $r^{1/4}$
    surface brightness law with $R_{eff}=1$.  To better represent the
    cuspy density profile graphically, the density eigenfunctions are
    shown relative to the deprojected $r^{1/4}$ law.}
  \label{fig:devauc}
\end{center}
\end{figure}

The background galaxian profile need not have finite mass and may be
cuspy.  For example, a basis set tailored to the singular isothermal
sphere only requires one to specify appropriate boundary conditions.
Boundary conditions corresponding to a disturbance not felt by in the
singular core and at large radii are:
\begin{equation}
  \left.\cases{
      d\Psi(r)/dr = 0 & $l=0$ \cr
    \Psi(r) = 0 & $l\not=0$ \cr
    }\right\}_{r\rightarrow0}
\end{equation}
and
\begin{equation}
  \left.\cases{
    d\Psi(r)/dr = 0 & $l=0$ \cr
    (l+1)\Psi(r)/r + d\Psi(r)/dr = 0 & $l\not=0$ \cr
    }\right\}_{r=r_t}
\end{equation}
where $\Psi(r) = \Psi_o(r) u(r)$.  These same boundary conditions
apply to the $r^{1/4}$ profile above.  The $l=0$ boundary conditions
ensure that the potential-density pairs are asymptotic to the
spherical background at small and large radii.  The $l\not=0$ boundary
condition at small radius is the standard zero potential that ensures
a single valued function.  At large radius, we choose the condition
obtained for an outer multipole. The four lowest-order $l=0$ pairs are
shown in Figure \ref{fig:singiso}.  The density functions are
multiplied by $r^2 \propto 1/\rho_o$ and, again, the lowest order
relative density function is constant as expected.

In addition, the background galaxian profile need not have an analytic
form.  For example, the spherically symmetric profile that results in
the empirical $r^{1/4}$ surface density law may be numerically
deprojected, tabulated and used as input to the SLE routines described
above.  A few of the lowest order potential-density pairs are shown in
Figure \ref{fig:devauc}.  The density functions are shown relative to
the background density.  Notice that the lowest order relative density
function is constant as expected.

\subsection{Three-dimensional cylindrical solutions for disks}

\subsubsection{Method}

\label{sec:PCA}

For the cylindrical case, there are boundary conditions at $r=0$,
$r=r_t$ and $z=\pm z_t$.  Here the situation is a bit trickier: the
general solution requires matching outgoing boundary conditions in two
dimensions.  However as $r_t\rightarrow\infty$, the multipole
expansion implies that equation (\ref{eq:sphbc}) applies to lowest
order in $1/r$ with $l$ replaced by $m$.  This technical
simplification is strong motivation for adopting the radial domain
$r\in[0,\infty)$ as is done here.  Implicit in equations
(\ref{eq:laplace}) and (\ref{eq:poisson}) is a separation constant
chosen to give oscillatory functions $Z(z)$ appropriate for a region
of non-zero density.  The functions match the outgoing Laplace
solution at the outer boundary.  By choosing the outer boundary of the
`pill box' sufficiently large (e.g. greater than ten scale heights),
we obtain boundary conditions appropriate for the isolated disk.  The
vertical biorthogonal functions are then the sines and cosines of the
discrete Fourier transform but over a vertical domain with twice the
height of interest.  This ensures that that force from density images
do not affect potential (cf. Eastwood \& Brownrigg
1979\nocite{EaBr:79}).

Experimentation suggests that $2^6=64$ wave numbers are sufficient to
adequately resolve the vertical structure.  Separating real and
imaginary parts (or equivalently, sine and cosine terms), this demands
128 coefficients per radial basis function!  Although this
trigonometric basis does not look the underlying basis, we can find an
orthogonal transformation which rotates the basis into one which look
like the desired equilibrium.  We do this by an empirical orthogonal
function analysis which is equivalent to principal component analysis
(see Weinberg 1996\nocite{Wein:96} for details).  In short, let the
vector $\Psi_i=\{p_{ij}\}$ be the potential basis functions evaluated
at the position of the $i^{th}$ particle.  The symmetric matrix
$S_{\mu\nu} = {1\over N} \sum_{i=1}^N p_{i\mu} p_{i\nu}$ measures the
weight of the particle distribution on the original basis.  By
diagonalizing this matrix, we determine an orthogonal transformation
to a new basis.  The lowest order basis function---the one with the
largest eigenvalue---best represents the underlying point distribution
followed in eigenvalue ranking by next best, etc.  The first few
functions usually represent most of the weight and this allows us to
reduce the 128 coefficients to between two and six.

\begin{figure}[htb]
  \mbox{
    \mbox{\epsfxsize=3.0in\epsfbox{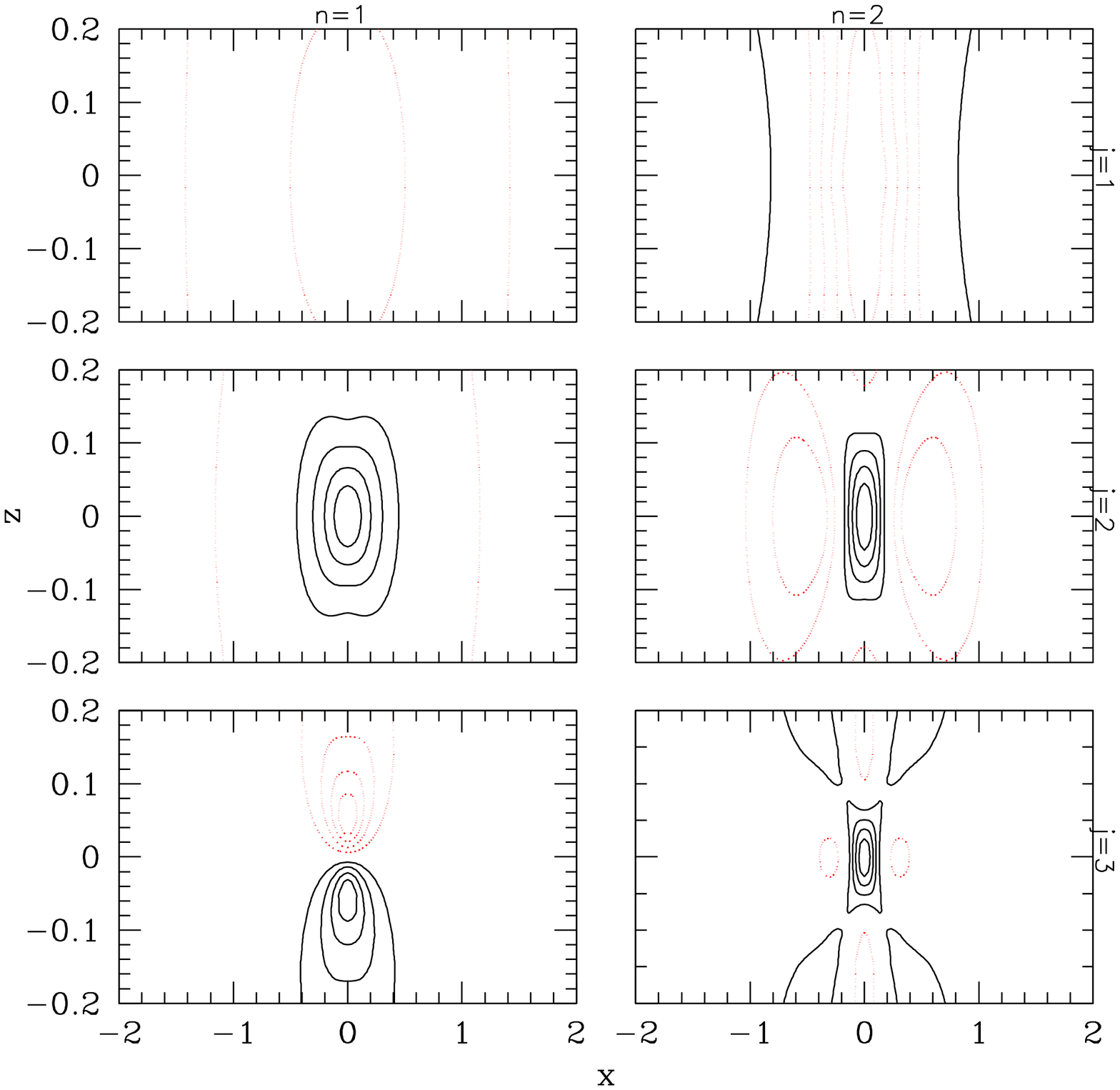}}
    \mbox{\epsfxsize=3.0in\epsfbox{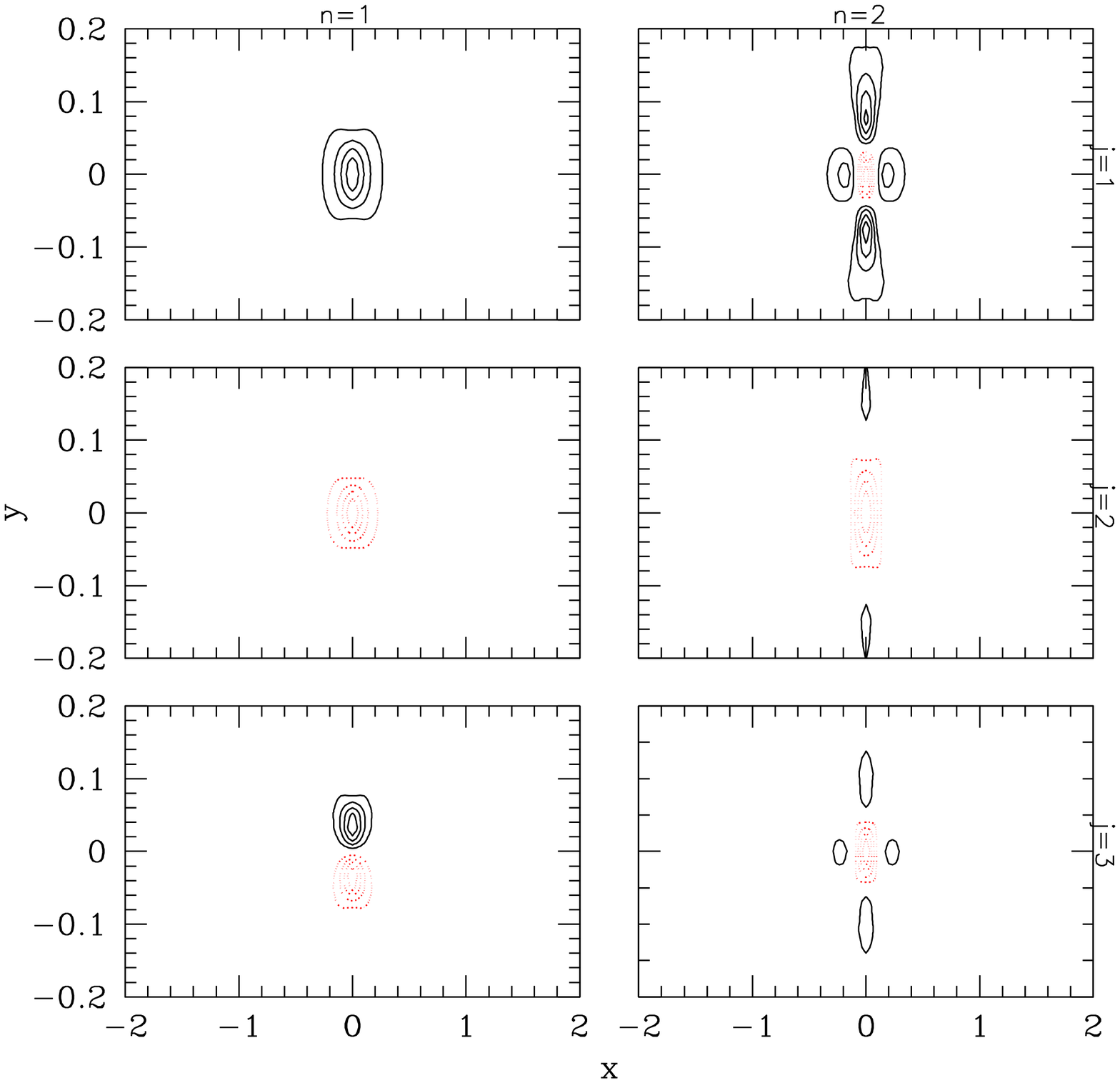}}
    }
  \caption{Six orthogonal potential and density pairs (left
    and right panels, resp.) labeled by vertical index $j$ and radial
    index $n$.  Azimuthal order is $m=0$.  Five contour levels are
    linearly spaced from from zero to the largest absolute peak value.
    Positive (negative) levels are shown as solid (dotted) lines.}
  \label{fig:eof6}
\end{figure}

Since the SLE solution is a good match to the radial profiles, we only
need the empirical transformation in the $z$-direction.  As an example
of these new functions, Figure \ref{fig:eof6} shows the first three
two-dimensional orthogonal functions for the two lowest radial orders
based on a Monte Carlo realization of the exponential disk with unit
scale length and scale height 1/10 using $10^5$ particles.  Following
the symmetry of the equilibrium model, the adaptive algorithm creates
the lowest order modes with even symmetry about the disk midplane.
However, the four or five lowest-order functions represent enough of
the odd component to follow the evolution (cf. Fig. \ref{fig:eof6}).

To summarize, the algorithm for the n-body force calculation for the
three-dimensional cylindrical basis is then as follows:
\begin{enumerate}
\item Compute $S_{\mu\nu}$ from particle distribution using the basis
  derived from equation (\ref{eq:poisson2}) with $Z(z)$ chosen as
  discussed above.
\item Compute transformation to new basis by solving for the
  eigenvectors.
\item Retain eigenvectors corresponding to the $M$ largest
  eigenvalues.  The value of $M$ may either be predetermined or
  computed adaptively from the cumulative distribution of
  eigenvalues (see Weinberg 1996 for details).
\item Tabulate the new orthogonal set and use this to evaluate force
  for some time-interval on order of a dynamical time for the problem
  of interest.
\item Goto 1.
\end{enumerate}
The computational bottleneck in this procedure is the construction of
$S_{\mu\nu}$ and computing Steps 1--3 can be a significant fraction of
the integrated time to advance the particles using the tabulated
orthogonal functions for several hundred time steps (30\% of the total
for the case illustrated in Fig. \ref{fig:eof6}).  Nonetheless, the
overall force evaluation is still very fast compared to other methods.

Although the underlying trigonometric basis is bounded vertically from
above and below, the boundary can be chosen large enough to permit
arbitrarily large vertical distortions.  Large vertical boundaries
require more wavenumber to achieve a fixed resolution.  In turn, more
wavenumbers affect the computational overhead in computing the
empirical basis but do not add to the CPU time required for the force
evaluation itself.  Therefore, large vertical boundaries remain
practical as long as the transform to the empirical basis described in
the algorithm above can be done infrequently.

\subsubsection{Examples}

\begin{figure}[htb]
\begin{center}
\mbox{
  \mbox{\epsfxsize=3.0in\epsfbox{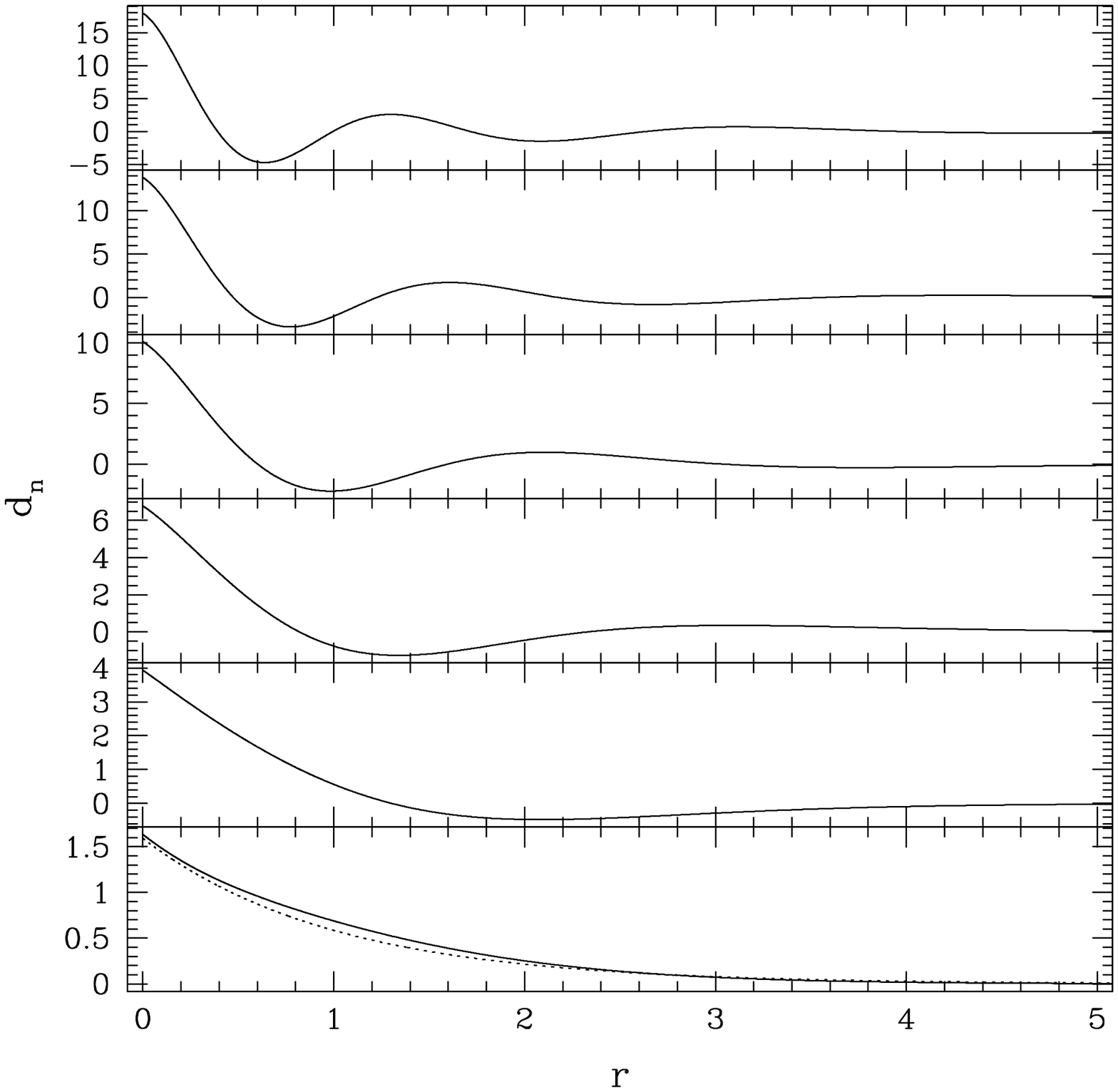}}
  \mbox{\epsfxsize=3.0in\epsfbox{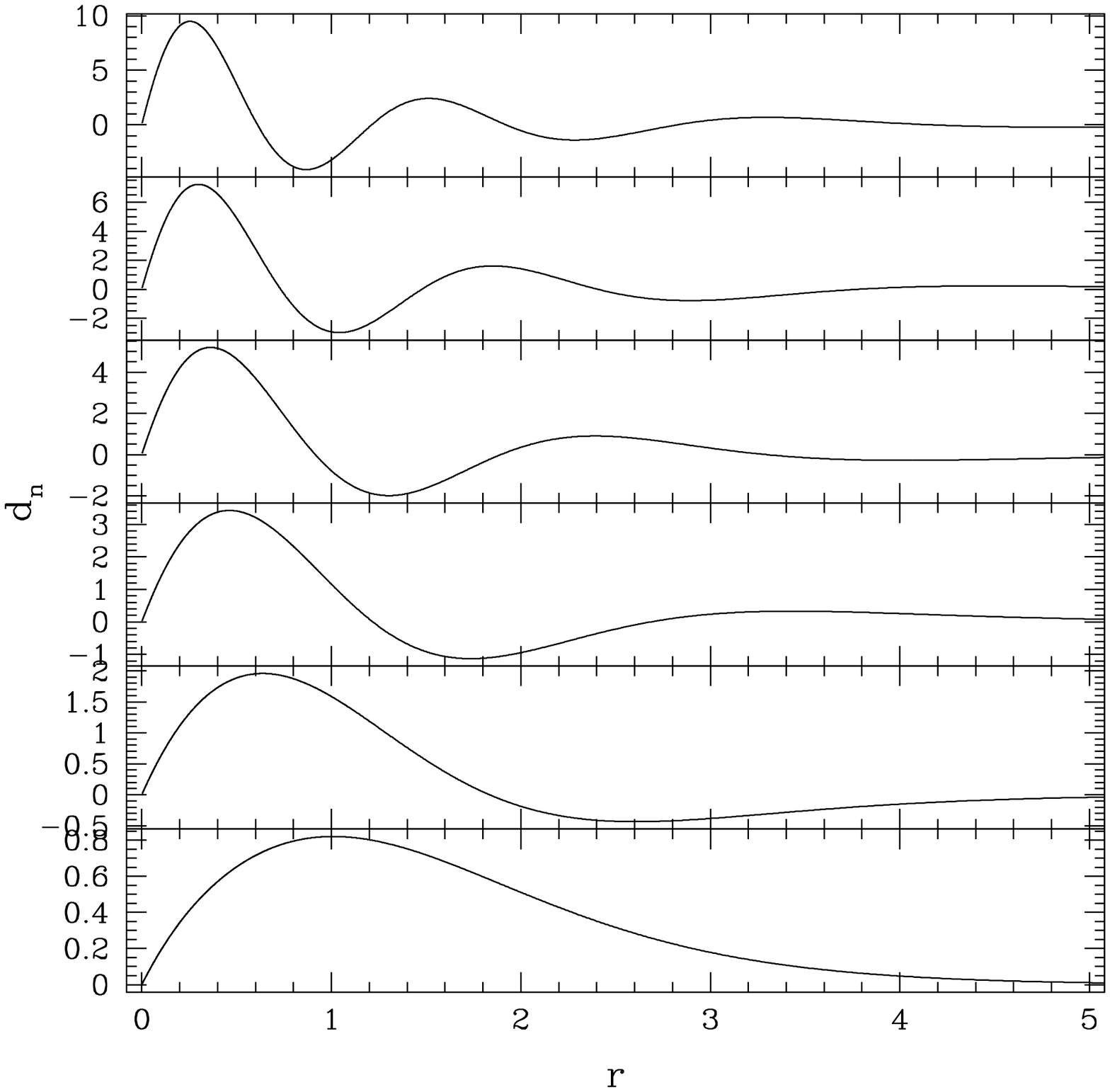}}
  }
\caption{First five density functions for $m=0$ (left) and $m=1$ (right) with
  $k=2\pi/5$ ordered from bottom to top.  The dotted curve on the
  lower-left-most panel shows the background exponential disk for
  comparison.}
\label{fig:dexp}
\end{center}
\end{figure}

\begin{figure}[htb]
\begin{center}
  \mbox{\epsfxsize=5.0in\epsfbox{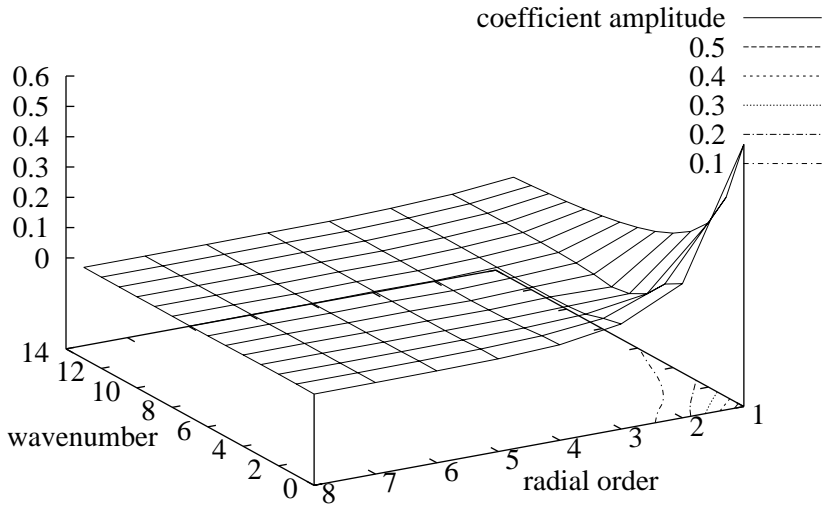}}
  \caption{Expansion coefficient amplitudes for an exponential disk
    with $\sech^2(z)$ vertical profile as a function of radial order
    and vertical wavenumber.}
  \label{fig:coefs}
\end{center}
\end{figure}

Here we build a basis set that closely matches the typical exponential
disk profile.  As described in \S\ref{sec:deriv}, we adopt an
axisymmetric separable density profile, $\rho(r, z) =
\rho_r(r)\rho_z(z)$, chosen to match the background, $\rho_r(r) =
\rho_o(r)$.  For this test, $\Psi_o\equiv-1/\sqrt{1+(r/a)^2}$ takes
care of the boundary conditions.  Recall that $\Psi_o$ and $\rho_o$
are not required to satisfy the Poisson equation; equation
(\ref{eq:poisson}) guarantees that the resulting basis functions will
be orthogonal regardless.  The results are shown in Figure
\ref{fig:dexp} for the four lowest radial terms for $m=0$ and $1$.
The exponential scale length $a=1$ and vertical boundary $L=10$ is
chosen to represent a disk with scale length to scale height ratio of
10:1.  The wavenumbers are $k=2\pi j/L j, j=0,1,\ldots, j_{max}$ for
pill box of half-height $L$.  The density functions in the figure have
$k=2\pi/5$ ($j=2$).  The lowest order $m=0$ case is compared with the
exponential disk (dotted).  For large $k$, the lowest order radial
function falls off more rapidly than the exponential disk.  However,
the series converges quickly in radial order and wavenumber as
demonstrated in Figure \ref{fig:coefs} which shows the coefficients
for an expansion of Monte Carlo realization of an exponential disk
with $\rho_z\propto\sech^2(z)$.  Good agreement demonstrates that
satisfactory results are obtained without exact Poisson solutions
$\rho_o$ and $\Phi_o$.  The biorthogonality condition (eq.
\ref{eq:ortho}) is good to one part in $10^9$.

The grid points for the Sturm-Liouville solution described in
\S\ref{sec:numerical} are chosen by mapping the semi-infinite interval
to the segment $[-1,1]$ using $x=(r-1)/(r+1)$ and choosing points
evenly spaced in $x$.  The Pruess \& Fulton algorithm can estimate the
grid automatically to optimize accuracy but this mapping provided
sufficiently high-accuracy and rapid execution.

I checked accuracy and consistency of the final basis set by
evaluating the gravitational force for a Monte Carlo distribution of
$10^5$ bodies with the proposed method and with a direct summation.
Contours of constant force are better than $1\%$ except where the
direct summation evaluation is badly affected by discreteness noise.


\section{Summary and conclusions}

This paper presents a numerical algorithm for constructing biorthogonal
expansion bases for use in n-body force calculation and linear
perturbation theory and explores its performance.  The major results
of this investigation are as follows:
\begin{enumerate}
\item This algorithm removes one of the remaining limitations of the
  self-consistent field (SCF) method by providing basis sets tailored
  to any background galaxian profile.
\item The algorithm is applicable to any separable coordinate system.
  This paper details and benchmarks its implementation for spherical
  and three-dimensional cylindrical bases.
\item Sturm-Liouville equation solvers are publically available (e.g.
  see Pruess \& Fulton 1993 for Fortran code) and a desired basis is
  readily obtained using equations (\ref{eq:sle1})--(\ref{eq:sle3}).
\item The main limitation of this method for n-body codes is the
  necessity to tabulate the basis functions rather than derive them
  from recursion relation on the fly (as in Clutton-Brock 1973
  and Hernquist \& Ostriker 1992).  On the other hand, this is largely
  a programming inconvenience; the algorithm is still easily
  parallelized and the table lookup is not a computational bottleneck.
\item For spherical expansions, the algorithm is conceptually
  equivalent to and computationally competitive with the published SCF
  expansions.  For three-dimensional cylindrical expansions, the
  coupling of the vertical and radial parts of the potential-density
  pairs requires an additional orthogonalization step.  This increases
  the computational overhead by up to 50\% but does not effect scaling
  with particle number or parallelizability.
\item The use of these basis sets is not limited to n-body simulation.
  They are easily used in semi-analytic linear perturbation
  calculations and, moreover, facilitate the comparison between the
  n-body and perturbation theory.
\end{enumerate}

\acknowledgements

I thank Lars Hernquiust and Neal Katz for discussion and suggestions.
This work was supported in part by NSF grant\# AST-9529328 and the
Alfred P. Sloan Foundation.

\end{document}